# Large-amplitude coherent spin waves exited by spin-polarized current in nanoscale spin valves

I. N. Krivorotov

*Department of Physics and Astronomy, University of California, Irvine, CA 92697-4575*

D. V. Berkov, N. L. Gorn

*Innovent Technology Development, Pruessingstraße 27B, D-07745 Jena, Germany*

N. C. Emley, J. C. Sankey, D. C. Ralph, and R. A. Buhrman

*Cornell University, Ithaca, New York 14853-2501*

**Abstract:**

We present spectral measurements of spin-wave excitations driven by direct spin-polarized current in the free layer of nanoscale $Ir_{20}Mn_{80}/Ni_{80}Fe_{20}/Cu/Ni_{80}Fe_{20}$ spin valves. The measurements reveal that large-amplitude coherent spin wave modes are excited over a wide range of bias current. The frequency of these excitations exhibits a series of jumps as a function of current due to transitions between different localized nonlinear spin wave modes of the $Ni_{80}Fe_{20}$ nanomagnet. We find that micromagnetic simulations employing the Landau-Lifshitz-Gilbert equation of motion augmented by the Slonczewski spin torque term (LLGS) accurately describe the frequency of the current-driven excitations including the mode transition behavior. However LLGS simulations give *qualitatively* incorrect predictions for the amplitude of excited spin waves as a function of current.

## I. INTRODUCTION

The recent discovery of persistent current-driven excitations of magnetization in magnetic nanostructures [1-13] creates new opportunities for studies of magnetization dynamics in extremely nonlinear regimes inaccessible with conventional techniques such as ferromagnetic resonance (FMR). It was recently demonstrated [14] that a spin-polarized current can excite motion of magnetization in metallic nanomagnets with precession cone angles over 30° - values far exceeding those achievable in typical FMR experiments performed on bulk and thin-film samples. There are two reasons why it is possible to have such large-amplitude current-driven motions of magnetization in nanomagnets: (i) suppression of Suhl instability processes [15,16] due to quantization of the magnon spectrum in the nanomagnet [17-28], and (ii) efficient amplification of spin waves by spin transfer torque that can act approximately as negative magnetic damping [1, 2].

The possibility of exciting large-amplitude oscillations of magnetization in magnetic nanostructures by spin-polarized current provides a unique testing ground for theories of nonlinear magnetization dynamics in ferromagnetic metals [29-32]. Most importantly, it gives an opportunity to test the validity of the Landau-Lifshitz-Gilbert (LLG) equation for the description of large-amplitude motion of magnetization. The LLG equation is phenomenological in nature and thus its applicability must be tested in every new type of experimental situation. This equation has proved to be largely successful in the description of persistent small-angle magnetic excitations [33] and transient large-angle magnetization dynamics [34,35] in thin films of ferromagnetic metals (with some notable exceptions [36]). However, it is not known *a priori* that the LLG equation is suitable for the quantitative description of a persistent magnetization precession with very large amplitude. For example, the phenomenological Gilbert damping term parameterized by a single constant in the LLG equation may prove to be an approximation suitable for description of small-angle dynamics but not valid in general. Recently, large-angle persistent motion of magnetization was studied in thin films of $Ni_{80}Fe_{20}$ by time-resolved measurements and a large increase of apparent damping was observed in the nonlinear regime [37]. However, measurements of *intrinsic* damping in continuous ferromagnetic films are obscured by generation of parametrically excited spin waves which give rise to at least a large portion of the increased damping found in [37]. This generation of parametrically pumped spin waves is expected to be suppressed in nanoscale ferromagnets [23] and thus information on the amplitude dependence of intrinsic damping can, in principle, be accessed. A number of recent models predict non-trivial angular dependence of damping [38,39] and suggest how it may depend on the rate and amplitude of magnetization precession [40] in metallic magnetic nanostructures. These predictions remain largely untested primarily due to the difficulty of exciting persistent large-amplitude magnetization dynamics in nanomagnets.

In this work we report a detailed comparison of experimentally measured spectra of current-driven magnetization oscillations in elliptical Py (Py ≡ $Ni_{80}Fe_{20}$) nanoelements to the results of full-scale micromagnetic simulations for these structures, and thus test the validity of the micromagnetic LLG approach for the description of strongly nonlinear oscillations of magnetization in magnetic nanostructures. We find that although simulations based on LLG equations augmented by Slonczewski spin torque term [1] (LLGS equations) can successfully mimic many properties of current-driven magnetization dynamics such as the current dependence of excitation *frequency* and abrupt frequency jumps with increasing current, they *qualitatively* fail to reproduce the dependence of the *amplitude* of current-driven spin waves as a function of current. Our results demonstrate the deficiencies of the current LLGS implementation for the description of spin-torque-driven magnetization dynamics and suggest the need for modification of this implementation for a quantitative description of large-amplitude magnetization motion. We suggest that it may be necessary to introduce a nonlinear dissipation or to consider effects of spin transfer from lateral spin diffusion that are not contained in our calculation.

## II. EXPERIMENT

*IIA. Sample Preparation and Characterization*

The current-perpendicular-to-plane (CPP) nanopillar spin valves for our experiments are prepared by magnetron sputtering of continuous magnetic multilayers onto an oxidized Si wafer followed by a multi-step nanofabrication process [7]. As a first step of the sample preparation process, a multilayer of Cu(80 nm)/ $Ir_{20}Mn_{80}$(8 nm)/ Py(4 nm)/Cu(8 nm)/ Py(4 nm)/ Cu(20 nm)/Pt(30 nm) is deposited onto a thermally oxidized Si (100) wafer by magnetron sputtering in a high vacuum system with a base pressure of 2 $10^{-8}$ Torr. The 80-nm Cu layer is used as the bottom electrode of the CPP spin valve. The Pt capping layer is employed for protection of the multilayer from oxidation during the nanopillar fabrication process. The multilayer is deposited at room temperature in a magnetic field of approximately 500 Oe applied in the plane of the sample and post-annealed at T = 250 °C for 80 minutes in the same field. We use a subtractive process employing e-beam lithography, photolithography and etching of the multilayer in order to define nanoscale spin valves of approximately elliptical shape with the major and minor axes of 130 nm and 60 nm, respectively, and with Cu electrodes making contact to the top and bottom of the spin valve as shown in Fig 1(a).

The role of the antiferromagnetic $Ir_{20}Mn_{80}$ layer in the spin valve structure is twofold: (i) to pin the direction of magnetization of the fixed Py nanomagnet at a non-zero angle with respect to the easy axis of the free nanomagnet using the exchange bias effect as shown in Fig. 1(b) and (ii) to suppress current-driven excitations of magnetization in the fixed nanomagnet due to the giant enhancement of Gilbert damping observed in exchange-biased ferromagnets [41,42].

The nominal direction of the exchange bias field set during the multilayer deposition and subsequent annealing is in the plane of the sample at 45° with respect to the major axis of the ellipse. However, within a set of forty samples we found significant (± 35°) sample-to-sample variations of the exchange bias direction, as determined from the Stoner-Wohlfarth (SW) fitting procedure described below. These sample-to-sample variations of the exchange bias direction are not surprising in a magnetic nanostructure and may be attributed to finite size effects [43-45] as well as to resetting of the exchange bias direction due to sample heating that occurs during lithography and ion milling process employed to define the nanopillar structure. In this paper we report experimental results for the most extensively studied sample although qualitatively similar results were obtained for other samples from the set of forty. The quantitative differences between the samples can be correlated with differences of the shapes of the hysteresis loop of resistance versus field such, as that shown in Fig. 1(c), and ultimately to variations of the direction of the exchange bias field. Samples with similar resistance versus field hysteresis loops exhibit similar spectral properties of the current driven magnetization oscillations. All measurements reported in this paper were made at $T = 4.2$ K.

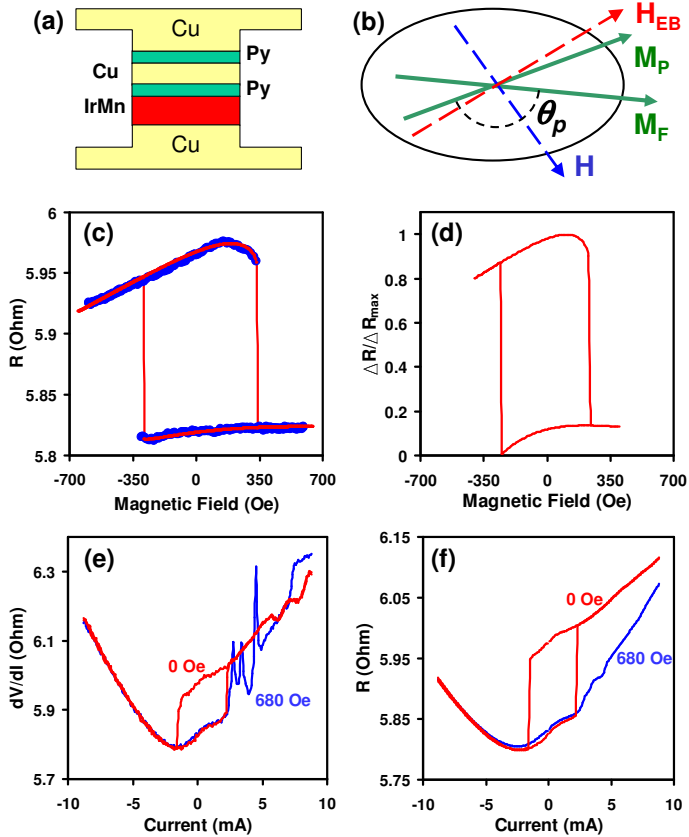

Figure 1. (Color online) (a) Schematic side view of the nanopillar spin valve used for studies of magnetization dynamics. (b) Schematic top view of the spin valve with approximate directions of magnetizations of the pinned, $\mathbf{M_P}$, and the free, $\mathbf{M_F}$, nanomagnets as well as the direction of positive external magnetic field, $\mathbf{H}$, and exchange bias field, $\mathbf{H_{EB}}$. $\theta_p$ is the equilibrium angle between $\mathbf{M_F}$ and the direction of spin torque applied to $\mathbf{M_F}$. (c) Experimentally measured resistance of the nanopillar as a function of the external magnetic field (circles) and a macrospin Stoner-Wohlfarth fit to the data (solid line) with the parameters described in text. (d) Resistance versus field obtained from micromagnetic simulations using the GMR asymmetry parameter $\chi = 0.5$, the exchange bias field magnitude $H_{EB} = 1600$ Oe and its direction $\theta_{EB} = 30°$ obtained from the macrospin fit shown in Fig. 1(c). (e) Differential resistance of the sample as a function of bias current measured at H = 0 Oe (red) and H = 680 Oe (blue). (f) DC resistance of the sample as a function of bias current obtained from the data in Fig. 1(e) by numerical integration.

We determine the direction and magnitude of the exchange bias field for each nanopillar sample by fitting the Stoner-Wohlfarth model to the experimental resistance-versus-magnetic-field hysteresis loop, such as that shown in Fig. 1(c). For the measurements reported in this paper, we apply the external magnetic field in the plane of the sample at 45° with respect to the ellipse major axis and approximately perpendicular to the exchange bias direction as shown in Fig. 1(b). Stoner-Wohlfarth simulations show that this choice of the bias field direction results in a weak dependence on the magnitude of external magnetic field for the equilibrium angle between magnetic moments of the free and pinned layers. According to Stoner-Wohlfarth simulations, the equilibrium angle between magnetic moments of the free and the pinned layer varies between 34° and 36° in the field range from 300 Oe to 1100 Oe. The solid line in Fig. 1(c) is a four-parameter Stoner-Wohlfarth fit to the data with the following fitting parameters: the exchange bias field magnitude, $H_{EB}$, its direction, $\theta_{EB}$, the magnetoresistance (MR) asymmetry, $\chi$, and the MR magnitude, $\Delta R$. The MR asymmetry parameter $\chi$ [46,47] describes a deviation of the angular dependence of the giant magneto-resistance (GMR) from a simple cosine form:

$$R(\theta) = R_0 + \Delta R \frac{1 - \cos^2(\theta/2)}{1 + \chi \cos^2(\theta/2)} \quad . \tag{1}$$

Here $\theta$ is the angle between magnetic moments of the pinned and the free layers. The Stoner-Wohlfarth fit shown in Fig. 1(c) yields $H_{EB} = 1.6 \pm 0.5$ kG, $\theta_{EB} = (30 \pm 6)°$, $\chi = 0.5 \pm 0.3$ and $\Delta R = 0.161 \pm 0.007$ Ohm. Two other parameters used in the Stoner-Wohlfarth simulations are the uniaxial shape anisotropy field, $H_K$, of the elliptical Py nanomagnets and the average dipolar coupling field between the fixed and the pinned layers, $H_{dip}$. The value of $H_K = 600$ Oe was obtained as the saturation field along the in-plane hard axis of the nanomagnet by employing micromagnetic simulations (OOMMF) [48]. The value of $H_{dip} = 80$ Oe was obtained by numerical integration of the dipolar coupling energy of the two uniformly magnetized Py nanomagnets. The value of $\chi$ obtained from our fitting procedure is significantly less than that reported for a similar structure in [47] ($\chi \approx 2$). The difference is probably due to the different values of the effective Py/Cu interfacial and Py bulk resistances in our spin valves, possibly due to inter-diffusion of metallic layers of the spin valve during the annealing process [49].

To test the validity of the Stoner-Wohlfarth approach for fitting the quasi-static MR hysteresis loop, we calculate the MR loop for this sample by employing full micromagnetic simulations [50] with the values of $H_{EB}$, $\theta_{EB}$ and $\chi$ obtained from the SW-fit. Other input parameters for micromagnetic simulations were obtained by direct measurements. The saturation magnetization $M_S$ of a 4-nm thick Py film sandwiched between two Cu films and subjected to the same heat treatment as the spin valves under study was measured by SQUID magnetometry and was found to be $M_S = 650$ emu/cm$^3$ at $T = 5$ K. The Gilbert damping parameter $\lambda = 0.025$ (needed for the dynamic simulations described in section III below) for these samples was measured by a pump-probe technique described in Ref. [14].

The result of the full-scale micromagnetic simulation is shown in Fig. 1(d). We find that the SW model is a reasonable approximation for the quasi-static hysteresis loop in that the coercivity predicted by micromagnetic simulation is ~ 80 % of that given by the SW model and the shapes of the Stoner-Wohlfarth and micromagnetic hysteresis loops are similar. However, we could not obtain a *quantitatively* correct fit of the measured GMR loop using full-scale micromagnetic simulations. We note that in full-scale simulations we do not have at our disposal the anisotropy field $H_K$ and the dipolar coupling $H_{dip}$ as adjustable parameters - the corresponding energy contributions are calculated from the material saturation magnetization and sample geometry. The discrepancy between the SW-simulations and full scale micromagnetic modeling means, first, that the fit parameters obtained from the SW-approximation should be considered not as exact values, but rather as reasonable guesses, and second, that some magnetic properties of the system under study (e.g., surface anisotropy, sample shape imperfections and the possible presence of antiferromagnetic oxides along the perimeter of the free layer nanomagnet [51]) are still not included in our model. Because the main goal of this paper is the study of *dynamic* system properties, we postpone the discussion of this quite interesting problem to future publications.

Figure 1(e) shows the measured differential resistance of the sample as a function of direct current flowing through the sample for $H = 0$ Oe and for $H = 680$ Oe. Positive current in this and subsequent figures corresponds to the flow of electrons from the free to the pinned layer. Figure 1(f) shows the DC sample resistance $R=V/I$ for $H = 0$ Oe and $H = 680$ Oe as a function of direct current, obtained by numerical integration of the differential resistance data in Fig. 1(e). The quasi-parabolic increase of the resistance with increasing current (most clearly seen for negative currents) is due to a combination of ohmic heating [52] and the Peltier effect in the nanopillar junction [53]. The other features in the plots of *dV/dI* versus *I* and *R*(*I*) such as hysteretic switching of resistance at $H = 0$ Oe or peaks in the differential resistance at $H = 680$ Oe are due to changes of magnetic state of the nanopillar. At fields below the coercive field of the free layer, we observe current-induced hysteretic switching between the low and the high resistance states. For fields exceeding the coercive field, the time-averaged resistance of the sample *R*(*I*) undergoes a transition from the low resistance state to an intermediate resistance state under the action of direct current as shown in Fig. 1(f) (e. g. $R(680 \text{ Oe}) = R(0 \text{ Oe}) - 0.27\Delta R$ for $I = 10$ mA). As we demonstrate below, this intermediate resistance state is a state of persistent current-driven magnetization dynamics for the free nanomagnet.

## *IIB. Measurements of current-driven oscillations of magnetization*

To measure the current-driven excitations of magnetization directly, we employ a spectroscopic technique developed in Ref. 10. Figure 2(a) schematically shows the measurement setup employed for detection of the current-driven excitations of magnetization. In this setup, direct current flowing perpendicular to the layers of the spin valve sample excites coherent spin wave modes in the free Py nanomagnet. Coherent spin waves give rise to a periodic variation of resistance of the spin valve due to the GMR effect, *R(t)*. Since the sample is current-biased ($I_{DC}$), periodic changes of resistance result in ac voltage generated by the device, $V(t) = I_{DC}R(t)$. This ac voltage is amplified with a microwave signal amplifier and its spectral content is recorded with a spectrum analyzer. A spectrum measured at zero dc bias current is subtracted from all spectra in order to eliminate a small background due to thermal and electronics noise.

Figures 2(b) and 2(d) show representative examples of typical spectra generated by the spin valve under direct current bias. The signals shown in Fig. 2(b) and 2(d) are normalized rms amplitude spectral density, *S*(*f*), defined below. This quantity characterizes the amplitude, frequency and coherence of oscillations of magnetization. To calculate *S*(*f*), we start with the power spectral density measured with the spectrum analyzer, $P_{an}(f)$. This quantity is corrected for frequency-dependent amplification and attenuation in the circuit between the spectrum analyzer and the nanopillar sample in order to obtain the power spectral density *P*(*f*) of the signal emitted by the sample into a 50-Ohm transmission line. This latter quantity is used to calculate the rms voltage spectral density *V*(*f*) of the GMR signal due to oscillations of magnetization at the nanopillar as $V(f) = (R_s + R_0)\sqrt{P(f)/R_0}$ [54]. In this expression, $R_0 = 50$ Ohm is the characteristic impedance of all components of the microwave circuit shown in Fig. 2(a) except for the nanopillar itself, and $R_S$ = 26 Ohm is the resistance of the nanopillar junction and leads. We define the normalized rms amplitude spectral density, $S(f)$ as the rms voltage spectral density $V(f)$ divided by the maximum rms GMR voltage signal amplitude $\langle V_{max} \rangle = \sqrt{\langle [(I\Delta R/2)\sin(\omega t)]^2 \rangle / 2} = I\Delta R / 2\sqrt{2}$ (where $\omega = 2\pi f$) achievable due to 360° uniform rotation of magnetization in the sample plane at a given current bias:

$$S(f) = \frac{V(f)}{\langle V_{max} \rangle} = \sqrt{8}\frac{V(f)}{I\Delta R}. \qquad (2)$$

The dimensionless integrated signal amplitude, $S_{int}$

$$S_{int} = \sqrt{\int_0^\infty S(f)^2 df} \qquad (3)$$

reaches its maximum value $S_{int} = 1/\sqrt{2}$ for the maximum possible GMR voltage signal due to 360° uniform rotation of magnetization in the sample plane:

$$V_{max}(t) = \frac{I\Delta R}{2}\sin(2\pi \cdot f \cdot t) \qquad (4)$$

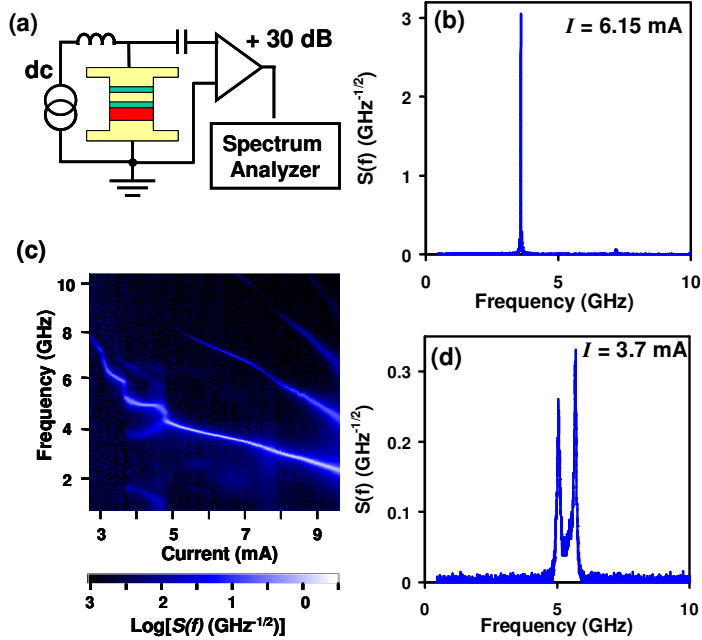

Figure 2. (Color online) (a) Circuit schematic for measurements of magnetization dynamics driven by a direct current. (b) Normalized rms-amplitude spectral density, $S(f)$, (defined in text) generated by the spin valve under a dc bias of 6.15 mA. (c) $S(f)$ as a function of current for the nanopillar spin valve measured at H = 680 Oe. (d) $S(f)$ at 3.7 mA and H = 680 Oe.

The integrated signal amplitude $S_{int}$ is a convenient dimensionless scalar quantity that characterizes the amplitude of magnetization precession. Its square is directly proportional to the integrated power emitted by the device. This dimensionless quantity is also convenient for comparison of experimental data to the results of micromagnetic simulations.

A typical experimentally measured spectrum $S(f)$ for our samples is characterized by a single frequency (the fundamental peak and higher harmonics such as that shown in Fig. 2(b)). However, for some values of the bias current, two peaks that are not harmonically related to each other are observed (Fig. 2(d)).

Figure 2(c) shows a summary of spectra generated by the sample as a function of the direct current bias, $I_{DC}$, measured at a fixed value of the applied magnetic field $H = 680$ Oe. The most important features of these data are:

1. The frequency of the current-driven excitations decreases with increasing current. This decrease of frequency with increasing current can be explained as a nonlinear effect arising from the dependence of the frequency of precessing magnetization on the precession amplitude [10,29,30,55,56].

2. The frequency of the current-driven excitations exhibits *two* downward jumps at $I \approx 3.7$ mA and 4.85 mA. The current values at which the frequency jumps occur coincide with the positions of the peaks in the plot of differential resistance versus current (Fig. 1(e)). A double-peak structure in the spectrum such as that shown in Fig. 2(d) is observed only for currents near frequency jumps, indicating that the apparent jumps are in fact non-hysteretic crossovers between two excitations with different frequencies. As the current is increased across the transition region, the emitted power is gradually transferred from the excitation with the higher frequency to the excitation with the lower frequency. We also observe that the linewidths of the current-driven excitations increase in the current intervals where two excitations coexist (e. g. compare Fig. 2(b) and 2(d)). In the current intervals where a single large-amplitude mode is excited, spectral lines as narrow as 10 MHz are observed while for currents in the mode transition regions spectral lines as wide as 250 MHz are found. The increase of the linewidth of the excitation indicates the decrease of its phase coherence [57-59]. The linewidth increase is observed in all transition regions suggesting that the decrease of coherence of the current-driven spin waves is induced by interaction between the two excited spin wave modes.

3. Modes with very low power visible only on the logarithmic amplitude scale of Fig. 2(c) are observed for currents above 3.7 mA. These modes are not harmonically related to the dominant modes. Although these modes emit low integrated power, they may play an important role in determining the coherence of the dominant spin wave excitations [60].

### III. NUMERICAL SIMULATIONS
### IIIA. *Methodological aspects*

Full-scale micromagnetic simulations of the current-induced magnetization dynamics in the nanopillar described above were performed using the commercially available simulation package MicroMagus [50] supplemented by a spin injection module. In this package, the magnetization dynamics are simulated by solving the stochastic LLG equation of motion for the magnetization $\mathbf{M}_i$ of each discretization cell in the form:

$$\frac{d\mathbf{M}_i}{dt} = -\gamma \cdot [\mathbf{M}_i \times (\mathbf{H}_i^{\text{det}} + \mathbf{H}_i^{\text{fl}})] - \gamma \cdot \frac{\lambda}{M_S} \cdot [\mathbf{M}_i \times [\mathbf{M}_i \times (\mathbf{H}_i^{\text{det}} + \mathbf{H}_i^{\text{fl}})]] \qquad (5)$$

Here the precession constant is $\gamma = \gamma_0/(1+\lambda^2)$, where $\gamma_0$ (> 0) is the absolute value of the gyromagnetic ratio. Our reduced dissipation constant $\lambda$ is equal to the dissipation constant $\alpha$ in the Landau-Lifshitz-Gilbert equation of motion written in the form $\dot{\mathbf{M}} = -\gamma_0 [\mathbf{M} \times \mathbf{H}] + (\alpha / M_S) \cdot [\mathbf{M} \times \dot{\mathbf{M}}]$. The *deterministic* effective field $\mathbf{H}_i^{\text{det}}$ acting on the magnetization of the *i*-th cell includes all standard micromagnetic contributions (external, anisotropy, exchange and magnetodipolar interaction fields) and the spin torque effect as explained below.

The random *fluctuation* field $\mathbf{H}_i^{\text{fl}}$ represents the influence of thermal fluctuations and has standard $\delta$-functional spatial and temporal correlation properties:

$$\langle H_{\xi,i}^{\text{fl}} \rangle = 0, \quad \langle H_{\xi,i}^{\text{fl}}(0) \cdot H_{\psi,j}^{\text{fl}}(t) \rangle = 2D \cdot \delta(t) \cdot \delta_{ij} \cdot \delta_{\xi\psi} \tag{6}$$

(*i, j* are the discretization cell indices; $\xi, \psi = x, y, z$) with the noise power $D$ proportional to the system temperature $D = \lambda/(1+\lambda^2) \cdot (kT/\gamma\mu)$; here $\mu$ denotes the magnetic moment magnitude for a single discretization cell. The justification for using $\delta$-correlated random noise for a finite-element version of an initially continuous system with interactions can be found, e.g., in [61].

The spin torque is taken into account adding the term $\Gamma = -f_J(\theta)[\mathbf{M} \times [\mathbf{M} \times \mathbf{p}]]$ to the equation of motion in the Gilbert form (see, e.g., [55]):

$$\frac{d\mathbf{M}}{dt} = -\gamma_0 [\mathbf{M} \times \mathbf{H}^{\text{eff}}] + \frac{\alpha}{M_S}[\mathbf{M} \times \dot{\mathbf{M}}] - \gamma_0 \cdot f_J(\theta) \cdot [\mathbf{M} \times [\mathbf{M} \times \mathbf{p}]] \tag{7}$$

where the *dimensionless* spin-torque amplitude $f_J$ depends on the angle $\theta$ between the magnetization $\mathbf{M}$ and the *unit vector* $\mathbf{p}$ of the polarization direction of the electron magnetic moments (in the spin-polarized current). From the computational point of view, this additional torque can be put into the effective field as $\mathbf{H}_{\text{ST}}^{\text{eff}} = \mathbf{H}^{\text{eff}} + f_J \cdot [\mathbf{M} \times \mathbf{p}]$, after which the equation (7) can be converted to the numerically more convenient form (5) in a standard way.

We use, in general, the asymmetric angular dependence of the spin torque amplitude $f_J(\theta)$ given by [46,62]:

$$f_J(\theta) = a_J \cdot \frac{2\Lambda^2}{(\Lambda^2 + 1) + (\Lambda^2 - 1)\cos\theta} \tag{8}$$

Here $a_J$ gives the (constant) value of the spin torque amplitude for a symmetric torque ($\Lambda = 1$); the asymmetry parameter $\Lambda$ can in principle be computed when the device configuration and various transport coefficients are known and is related to the GMR asymmetry parameter $\chi$ in Eq. (1) via $\Lambda^2 = 1 + \chi$. The expression (8) is only strictly valid for symmetrical spin valves [46, 62] with identical ferromagnetic layers and identical top and bottom leads. Expression for $f_J(\theta)$ in an asymmetric device is more complex [62] and involves effective resistances of the ferromagnetic layers and leads. However, we use a simplified expression (8) for $f_J(\theta)$ in our simulations for three reasons. First, we do not expect the spin torque asymmetry of our device (with respect to the above-mentioned effective resistances) to be large because the thicknesses of two ferromagnetic layers of the spin valve are identical and the thicknesses of non-magnetic leads are not very different. Second, the spin diffusion length of Py (~ 5 nm [63]) is similar to the thickness of Py layers in our spin valve structure, which substantially decreases the influence of the transport properties of the leads on spin torque [62]. Third, Eq. (8) can be considered as the *simplest* form (apart from the form with $f_J = Const$) for studying the effect of the asymmetry of the spin torque angular dependence on the magnetization dynamics. The more complex expression derived in [62] can be investigated after the effect of the simplest form of spin torque asymmetry given by Eq. (8) is understood.

The remaining simulation methodology is similar to that described in [64]. We simulate spin-torque-driven excitations in the free ferromagnetic layer only. We neglect, magnetostatic and RKKY interactions between the free and pinned (AF-coupled) Py layers. This approximation is justified because, first, the RKKY exchange coupling via the thick ($h_{sp}$ = 8 nm) Cu spacer is negligibly small and, second, the dipolar field acting on the free layer from the fixed one is on average (~ 80 Oe) much smaller than the external field. The free layer (130 x 60 x 4 $nm^3$ ellipse) is discretized into 50 x 24 x 1 cells; we checked that further refinement of the grid did not lead to any significant changes in the results.

The magnetic parameters of the free Py layer used in simulations are: saturation magnetization $M_S$ = 650 emu/$cm^3$ (measured by SQUID magnetometry as explained in Section II); exchange constant A = 1.3 x $10^{-6}$ erg/cm (standard value for Py); the random magnetocrystalline anisotropy was neglected due to its low value ($K_{cub}$ = 5 x $10^3$ erg/$cm^3$) for Py. The dissipation parameter is set to $\lambda$ = 0.025 (see also Section II).

As it will be demonstrated below, variations in the direction of the spin current polarization vector **p** (opposite to the magnetization of the pinned layer, $\mathbf{M_P}$) in Eq. (7) can result in qualitative changes of magnetization dynamics and thus the orientation of **p** is a very important parameter of the problem under study. As explained in Section II, the direction of **p** could not be determined quantitatively from the available experimental data. To understand the dependence of the magnetization dynamics on the orientation of **p**, we first study magnetization dynamics for **p** directed opposite to the exchange bias field extracted from the GMR hysteresis fit as described in Section II (i.e., the angle between **p** and the positive direction of the *x*-axis was set to $\theta_\mathbf{p} = 150°$). Then we perform two additional simulation sets for larger ($\theta_\mathbf{p} = 170°$) and smaller ($\theta_\mathbf{p} = 130°$) values of the equilibrium angle between magnetization and current polarization to study the effect of the spin polarization direction on the current-driven dynamics.

The computation of the Oersted field $\mathbf{H}_{Oe}$ induced by the current flowing through the spin valve is also a non-trivial issue. In principle, its precise evaluation requires the exact knowledge of the 3D current distribution in the device itself and especially in adjacent electrical contact layers, which is normally not available. For this reason the Oersted field is usually computed assuming that the current is distributed homogeneously across the nanopillar cross-section. Further, one of the following approximations is used: (i) one assumes that $\mathbf{H}_{Oe}$ is created by the *infinitely long* wire with the cross-section corresponding to that of the nanopillar (in our case the ellipse with $l_a$ x $l_b$ = 130 x 60 $nm^2$); or (ii) the contribution to the Oersted field from the current *inside the nanopillar itself only* is included (i.e., $\mathbf{H}_{Oe}$ is created by the piece of the wire with the length equal to the nanopillar height $h_{tot}$). Both approximations deliver the same result for nanopillars with the height much larger than their characteristic cross-section size ($h_{tot}$ >> max($l_a,l_b$)), which is, however, very rarely the case for experiments performed up to now in the nanopillar geometry. In particular, in our situation the opposite inequality is true ($l_a > h_{tot}$). Taking into account that the first approximation is also reasonably accurate for the system where the distribution of currents in the nanopillar and adjacent leads is axially symmetric, and that the geometry of the electric contacts in our device is also highly symmetric, we have chosen the first method to calculate $\mathbf{H}_{Oe}$. However, we point out once more that the influence of the Oersted field may be very important (see discussion of the results given below), so that more precise methods for its calculation are highly desirable.

The magnetization dynamics were simulated by integrating Eq. (5) with the spin-torque term included; the integration was performed using the Bulirsch-Stoer algorithm [65] with the adaptive step-size control additionally optimized. (The adaptive step-size control is especially important when the magnetization state significantly deviates from a homogeneous one.) For each current value (each value of $a_J$ in our formalism) the dependence of the magnetization on time for every discretization cell were saved for the physical time interval $\Delta t = 400$ ns. The spectral analysis of these magnetization 'trajectories' was performed using either (i) the Lomb algorithm (as described in [64]) especially designed for non-evenly spaced sequences of time moments as provided by the adaptive integration method or (ii) interpolation of the 'raw' results onto an evenly spaced temporal grid and usage of the standard FFT-routines. Results of both methods turned out to be equivalent within the statistical errors.

### IIIB. Simulation results

The decisive advantage of numerical simulations is the possibility to study and understand the influence of all relevant physical factors separately. For this reason we start from the 'minimal model', where the influence of the Oersted field and thermal fluctuations is neglected and the spin torque is assumed to be symmetric ($\Lambda = 1$ in Eq. (8)), and then switch on in succession all the factors listed above to analyze their influence on the magnetization dynamics.

*Minimal model*. Results for this model, for which the Oersted field and thermal fluctuations are not included and the spin torque amplitude does not depend on the angle between magnetization and current polarization, $f_J(\theta) = $ Const ($\Lambda = 1$), are presented in Fig. 3(a). For the spin orientation angle $\theta_p = 150°$ used for the first simulation series, the critical spin torque value for the oscillation onset was found to be $a_J^{cr} \approx 0.308(2)$. Using the simplest expression for the spin torque given, e.g., in [62], one can easily derive the relation between the reduced spin torque amplitude $a_J$ and other device parameters as

$$a_J = \frac{\hbar}{2} \cdot \frac{j}{|e|} \cdot \frac{1}{d} \cdot P \cdot \frac{1}{M_S^2} \quad , \tag{9}$$

where $e$ is the electron charge, $j$ is the electric current density, $d$ is the thickness of a magnetic layer subject to a spin torque, and $P$ is the degree of spin polarization of the electrical current. Using the definition of the current density $j = I/S_{elem}$ ($I$ is the total current and $S_{elem}$ is the area of the nanopillar cross-section) and substituting the values for the experimentally measured critical current $I_{cr} \approx 2.7$ mA and the threshold for the oscillation onset $a_J^{cr} \approx 0.3$ found in simulations, we obtain that the polarization degree of the electron magnetic moments is $P \approx 0.32$. From the relation between the critical current $I_{cr}$ and the critical spin torque amplitude $a_J^{cr}$, the proportionality factor $\kappa$ between the spin torque amplitude $a_J$ used in simulation and the experimental current strength $I$ (in mA) is $\kappa \approx 0.11$ (mA$^{-1}$) (whereby $a_J = \kappa I$) for this spin polarization direction ($\theta_p = 150^\circ$).

The simulated spectral lines are very narrow (mostly < 100 MHz) for all values of the spin torque amplitude $a_J^{cr} \leq a_J \leq 2.0$, which means that for this simplest model a transition to a quasichaotic regime similar to that found in [64] does not occur in the region of currents studied. For this reason we show in Fig. 3(a) only the positions of the spectral maxima of the $M_z$-component as a function of $a_J$ (red circles). In addition to the narrow lines, this minimal model also reproduces two other important qualitative features of the experimental results (see Fig. 2(c)): (i) a rapid decrease of the oscillation frequency with increasing current immediately after the oscillation onset and (ii) two downward frequency jumps at higher current values.

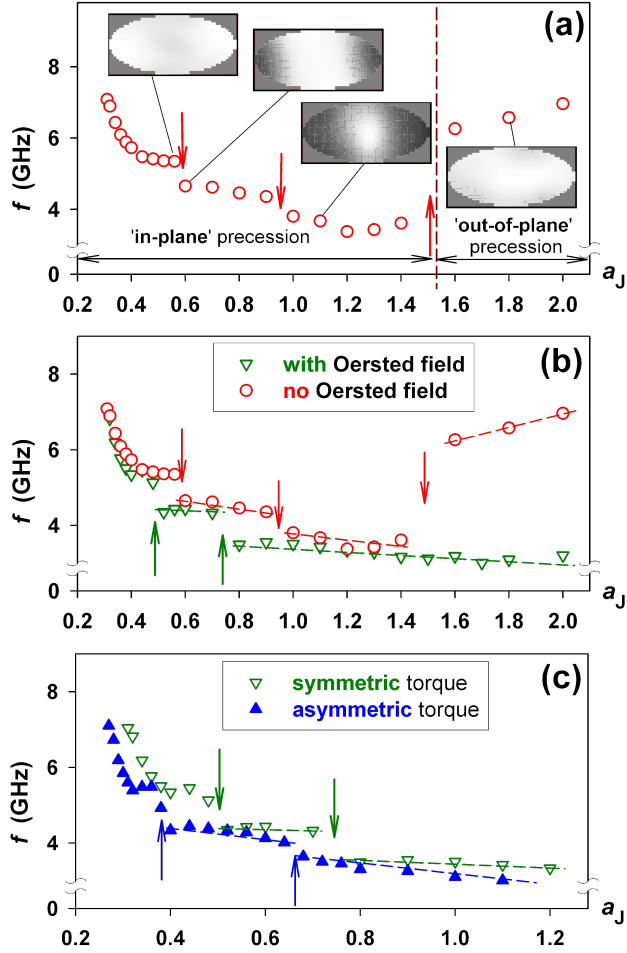

Figure 3. (Color online) (a) Dependence of the frequency of magnetization oscillations, $f$, on spin torque amplitude, $a_J$, calculated in the 'minimal model' (see text for details). Gray-scale maps represent spatial distributions of the oscillation power for chosen $a_J$-values (bright corresponds to maximal oscillation power). (b) Oersted field effect: $f(a_J)$ without (open circles) and with (open triangles) the Oersted field included in the simulations; arrows indicate the positions of frequency jumps, straight lines are guides to the eye. (c) Effect of the torque asymmetry: $f(a_J)$ for the symmetric ($\Lambda = 1.0$, open triangles) and asymmetric ($\Lambda^2 = 1.5$, solid triangles) spin torque.

The first feature, the rapid decrease of the oscillation frequency immediately after the oscillation onset, is a nonlinear effect due to the rapid growth of the oscillation amplitude with increasing current. In the nonlinear regime, the frequency decreases with increasing amplitude because the length of the precession orbit grows faster than the magnetization velocity. The corresponding effect was obtained analytically in [29, 30, 56] and observed numerically in our full-scale micromagnetic simulations [64] of the experiments published in [10].

The second important observation, the existence of two downward frequency jumps with increasing current, cannot yet be explained using analytical theories, and such jumps are absent in the macrospin description of current-driven magnetization dynamics. Spatially resolved spectral analysis of our simulation data reveals that these jumps correspond to transitions between strongly nonlinear oscillation modes (see spatial maps of the magnetization oscillation power in Fig. 3(a)). With each frequency jump, the mode becomes more localized but the oscillation power is still concentrated in one single-connected spatial region which has no node lines. We discuss these modes in more detail below when analyzing results for different current polarization directions. An analytical theory of the nonlinear eigenmodes of a resonator having the correct shape would be required to achieve a thorough understanding of this phenomenon.

In the minimal model we also observe for the current strength $a_J > 1.5$ the so-called 'out-of-plane' coherent precession regime for which the magnetization acquires a non-zero time-average component perpendicular to the sample plane. This regime is characterized by frequency *increasing* with current and is well known from analytical consideration and numerical simulations - [66, 67] and was experimentally observed for a nanopillar sample without exchange bias in [68]. However, for our system this type of mode is an artifact of the 'minimal model' (due to the absence of the Oersted field) and it was not observed experimentally.

*Effect of the Oersted field.* The effect of the Oersted field is demonstrated in Fig. 3(b) where the dependences of the oscillation frequency on the spin torque magnitude $a_J$ are shown without (red circles, identical to Fig. 3(a)) and with the Oersted field (green triangles). To compute the Oersted field, we have used the proportionality constant between the spin torque magnitude $a_J$ and experimental current value $I$ (in mA) assuming that the simulated threshold value $a_J^{cr} \approx 0.31$ corresponds to the experimentally measured critical current $I_{cr} \approx 2.7$ mA.

The results shown in Fig. 3(b) demonstrate that the Oersted field has two major effects on magnetization dynamics. First, this field eliminates the out-of-plane precession: inspection of magnetization trajectories shows that for all $a_J$ values they correspond to 'in-plane' steady-state oscillations. As a consequence, the Oersted field eliminates the upward frequency jump in the $f(a_J)$-dependence. The suppression of the out-of-plane mode occurs because the Oersted field is a strongly inhomogeneous *in-plane* field that keeps magnetization close to the plane of the sample.

The second effect of the Oersted field is a significant shift to lower values for the currents corresponding to the frequency jumps. This can be explained as follows: the Oersted field is highly inhomogeneous, with its maximal values at the edges of the elliptical element. For this reason it should suppress magnetization oscillations at the element edges, thus favoring spatial oscillation modes localized near the element center such as those shown in Fig. 3(a). Hence the transition from the homogeneous mode to more localized ones should occur for lower currents when the Oersted field is taken into account.

*Effect of the spin torque asymmetry.* It can be seen directly from Eq. (8) that for $\pi/2 < \theta_p < 3\pi/2$ the spin torque magnitude in the case of positive GMR asymmetry ($\chi > 0$, $\Lambda > 1$) is larger than for the symmetric ($\chi = 0$, $\Lambda = 1$) case. This difference is expected to result in a *decrease* of the steady-state precession frequency at a given current value because larger spin torque results in larger amplitude of magnetization oscillations. For the system studied in this paper, the GMR asymmetry is relatively low (the Stoner-Wohlfarth fit of the quasi-static GMR curve gave the value $\chi = 0.5$, and $\Lambda^2 = 1.5$), so that the expected frequency decrease is quite weak, but it is nevertheless clearly visible when comparing the $f(a_J)$-dependences for symmetric and asymmetric cases in Fig. 3(c). The frequency decrease is greatest in the low-current region where the dependence of the oscillation frequency on current is very steep. The asymmetric torque also leads to a small decrease of the threshold current for the oscillation onset, which becomes $a_J^{cr} \approx 0.27$.

A more important effect of introducing the spin torque asymmetry is a shift of the frequency values where the transitions between nonlinear eigenmodes of the system (accompanied by the frequency jumps as explained above) take place. Even for the relatively low value $\Lambda^2 = 1.5$ used in our simulations, this shift is significant (see Fig. 3(c)). The reason, again, is that the asymmetric torque form gives a larger spin torque magnitude for a given current value, so that the transitions to more-localized modes occur earlier.

*Influence of thermal fluctuations.* To take into account the influence of thermal fluctuations, we have first to estimate the real temperature of the sample. Although the experiments were performed at liquid helium temperature, $T = 4.2$ K, Joule heating of our multilayer nanoelement due to the direct current through the device was unavoidable [52]. To estimate the maximal temperature of the nanoelement, we have measured (i) the temperature dependence of its resistance $R_0(T)$ in the absence of any dc-current by heating the whole setup and (ii) the dependence of the resistance on current $R_0(I)$ measured at positive current and $H = 680$ Oe as shown in Fig. 1(f). The increase of the resistance with current is due both to the excitation of coherent magnetization oscillations and Ohmic heating, therefore an estimate of the sample temperature from $R_0(I)$ gives an *upper bound* on the temperature of the sample. Comparison of $R_0(T)$ and $R_0(I)$ shows that the nanoelement temperature does not exceed $T \approx 60$ K for the highest current $I = 10$ mA used in the measurements. Taking into account that this maximal temperature is relatively low, we have simply adopted a linear interpolation between the lowest temperature $T = 4$ K for $I = 0$ mA and $T \approx 60$ K for $I \approx 10$ mA (with $I$ converted into the reduced spin torque amplitude $a_J$) for our simulations.

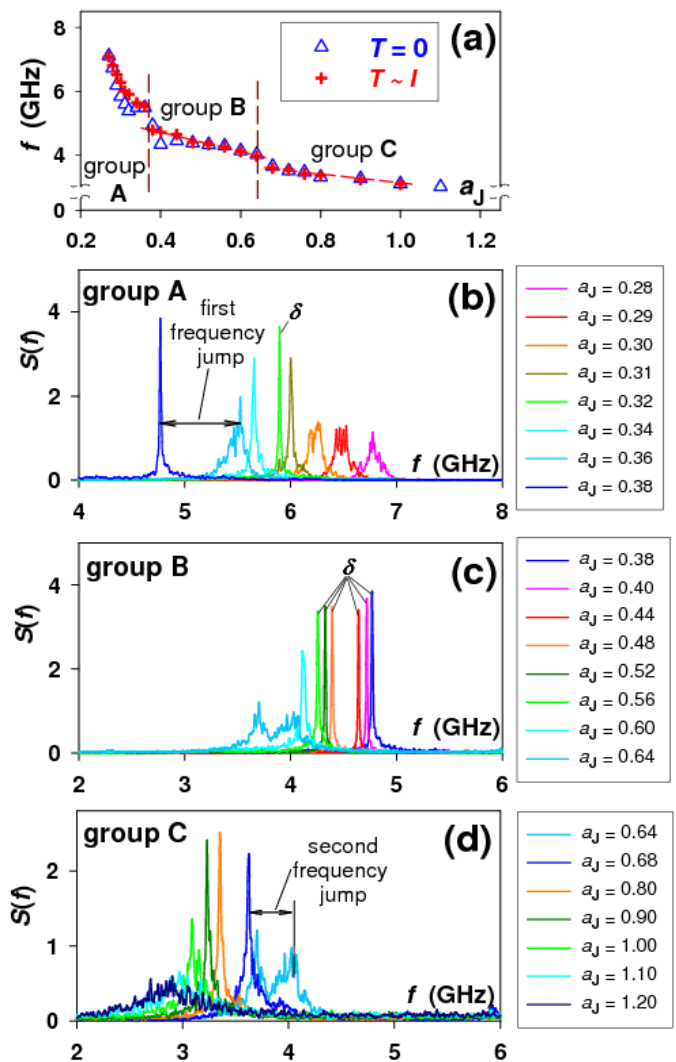

Figure 4. (Color online) (a) Effect of thermal fluctuations on the frequency of magnetization oscillations: $f(a_J)$ for $T = 0$ (triangles) and for $T \propto I \propto a_J$ (crosses). Simulated rms-amplitude spectral densities $S(f)$ for oscillations (b) before the first frequency jump, (c) between the first and the second frequency jump, and (d) after the second frequency jump. The linewidth of the peaks marked with δ is below the resolution limit of our numerical simulations. See the text for the detailed analysis of these spectra.

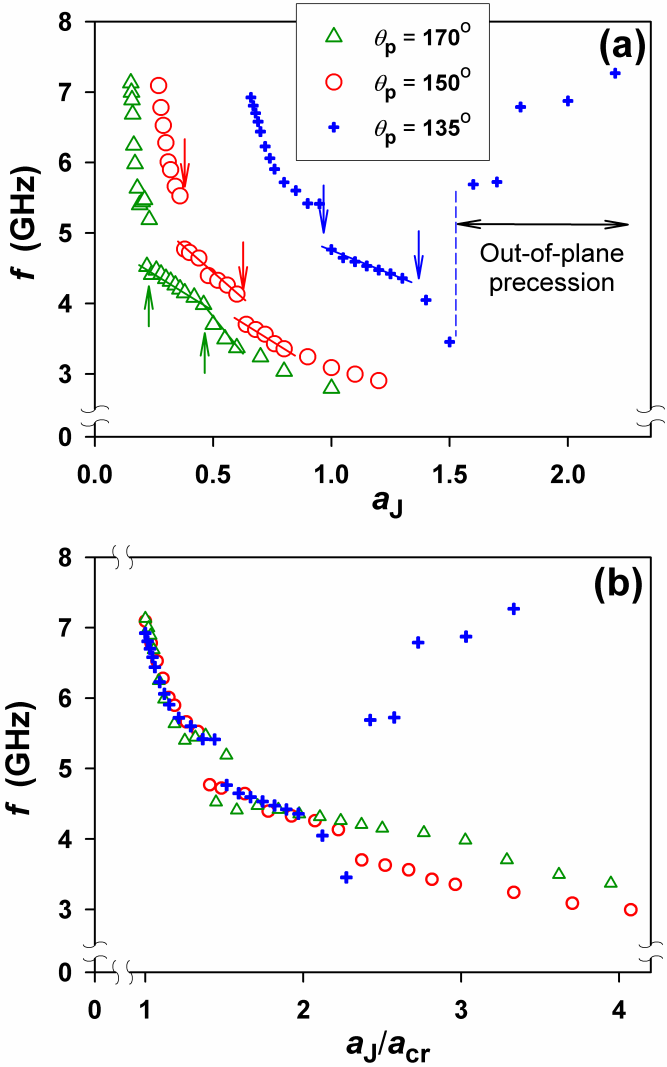

Figure 5. (Color online) (a) Dependence of the frequency of oscillations on spin torque amplitude, $a_J$, for different polarization angles of the spin current, $\theta_p$, as defined in Fig. 1(b). (b) The same frequencies plotted as functions of the normalized spin torque amplitude $a_J / a_J^{cr}$.

The dependences of the excitation frequency on current $f(a_J)$ for $T = 0$ and $T \propto a_J$ with the proportionality factor $\kappa$ calculated as explained above are compared in Fig. 4(a) (for both simulation sets the effect of the Oersted field and the spin torque asymmetry with $\Lambda^2 = 1.5$ are included). It is clear that due to the relatively low temperatures, thermal fluctuations have a minor effect both on the oscillation frequency and on the positions of the frequency jumps.

*Influence of the spin current polarization direction.* The polarization direction $\theta_p$ of the electron magnetic moments in the *dc*-current is expected to be one of the most important parameters of the problem. First, the onset threshold for oscillations should depend strongly on this polarization direction [55,69]. Second, the relative strength of the Oersted field (with respect to the spin torque magnitude $a_J$) also should depend on $\theta_p$, because the Oersted field for different $\theta_p$ is computed assuming that the threshold value $a_J^{cr}$ always corresponds to the experimentally measured critical current $I_{cr} \approx 2.7$ mA. Since $\theta_\mathbf{p}$ could not be accurately determined from the fit of the quasistatic MR hysteresis loop (see discussion in Section IIA), we have carried out additional series of simulation runs to study the effect of the spin current polarization direction on the magnetization dynamics.

The results of these simulations are summarized in Fig. 5, where we show the dependences of the oscillation frequency on the spin torque magnitude $f(a_J)$ (Fig. 5(a)) and on spin torque magnitude normalized by the threshold value $a_J^{cr}$ for the corresponding angle $f(a_J / a_J^{cr})$ (Fig. 5(b)).

The dependence $f(a_J)$ for $\theta_\mathbf{p} = 150°$, i.e., for the case which a detailed analysis has been presented above, is shown in this figure with open circles. For the *increased* polarization orientation angle $\theta_\mathbf{p} = 170°$ (open triangles) the onset threshold for the magnetization dynamics *decreases* from $a_J^{cr}(\theta_\mathbf{p} = 150°) \approx 0.27$ to $a_J^{cr}(\theta_\mathbf{p} = 170°) \approx 0.15$ in a qualitative agreement with the Slonczewski's prediction for the macrospin model ($I_{cr} \sim 1/|\cos\theta_\mathbf{p}|$) [1] experimentally confirmed in Ref. [69]. The first frequency jump with increasing current is still present, but instead of the second jump we observe a kink in the $f(a_J)$-curve (see Fig. 5(a)). We note that the importance of the Oersted field relative to the spin torque effect in this case is much larger than for $\theta_\mathbf{p} = 150°$ for the following reason. The Oersted field is always computed assuming that the critical value of the spin torque magnitude $a_J^{cr}$ corresponds to one and the same physical current value $I_{cr} \approx 2.7$ mA. This means, that if $a_J^{cr}$ decreases, the *same* Oersted field corresponds to *smaller* $a_J$-values so that the importance of the Oersted field effect increases relative to the spin torque action.

For a *smaller* polarization angle ($\theta_\mathbf{p} = 135°$, results are shown on Fig. 5 with crosses) the critical value of $a_J$ increases ($a_J^{cr}(\theta_\mathbf{p} = 135°) \approx 0.66$), so that the influence of the Oersted field is weaker than for $\theta_\mathbf{p} = 150°$. This leads, in particular, to the reappearance of the out-of-plane oscillation regime, which manifests itself in the increase of the oscillation frequency with increasing $a_J$. Recall that the out-of-plane precession regime was found for $\theta_\mathbf{p} = 150°$ in the absence of the Oersted field, but was suppressed by this field as explained above (see Fig. 2(b)). For the angle $\theta_\mathbf{p} = 135°$ the Oersted field is not strong enough to eliminate this regime when the spin torque magnitude increases.

To compare magnetization dynamics for various spin polarization angles we plot the frequency of oscillations for all three values of $\theta_\mathbf{p}$ studied as a function of spin torque magnitude normalized to its critical value, $a_J / a_J^{cr}$ (Fig. 5(b)). The most striking feature of the $f(a_J / a_J^{cr})$ curves for various angles $\theta_\mathbf{p}$ is that they all nearly collapse onto the universal $f(a_J / a_J^{cr})$ dependence for $a_J / a_J^{cr}$ values up to $a_J / a_J^{cr} \approx 2$. This region includes, in particular, the fast frequency decrease after the oscillation onset (see the discussion of this nonlinear effect above) and the first frequency jump arising for all spin polarization directions at almost the same value of $a_J / a_J^{cr} \approx 1.5$.

These results clearly demonstrate that the initial nonlinear rapid frequency decrease and the first frequency jump are universal for the system under study, whereas further behavior of the magnetization dynamics (in particular, the existence of the second frequency jump) are much more subtle features and thus may vary from sample to sample. The first frequency jump is always present because it marks the transition from the homogeneous to a localized oscillation mode (see Fig. 3(a)) which is always accompanied by an abrupt change of the oscillation frequency. The next frequency jump for the situation when the Oersted field is neglected corresponds to the transition between the modes with different (but symmetric) localization patterns - before the second jump the mode is localized in the direction along the major ellipse axis only, whereas after this jump the new mode is confined in both directions (compare second and third maps in Fig. 3(a)). This latter transition is strongly disturbed by the Oersted field, which leads, in particular, to strongly asymmetric spatial mode patterns for localized modes. This may eliminate the qualitative differences between modes with different localization patterns that give rise to the frequency jumps.

## IV. DISCUSSION

Having developed an understanding about how the various parameters influence the simulated dynamics, we can proceed with the analysis of magnetization oscillation spectra and a comparison with the experimentally observed magnetoresistance power spectra.

In Fig. 4(a) we display the dependence of the simulated spectral maxima frequencies on the spin torque amplitude $a_J$, in the presence of thermal fluctuations. These simulations take into account all the physical factors which are generally included in a state-of-the-art micromagnetic model. Spectral amplitudes of *magnetoresistance* oscillations are displayed in Fig. 4(b)-(d). The spectra can be divided into the following three groups: (i) from the oscillation onset to the first frequency jump (group A, Fig. 4(b)), (ii) from the first to the second frequency jump (group B, Fig. 4(c)) and after the second frequency jump (group C, Fig. 4(d)). For all groups, the frequency of the spectral maximum decreases monotonically with the spin torque magnitude $a_J$. The dependencies of the line width and the integrated spectral power (see Fig 6(b)) on the spin torque magnitude require special discussion.

For the first group - spectra from the oscillation onset to the first frequency jump - the line width for small $a_J$ is relatively large (~ 100 MHz) due to a relatively large influence of thermal fluctuations on small-amplitude motion of the magnetization [60]. The oscillation amplitude grows rapidly with increasing current (compare spectra for $a_J$ = 0.30, 0.31 and 0.32) and the linewidth strongly decreases (to ~ 20 MHz for $a_J$ = 0.32), which is due to an increasing contribution from the spin-torque driven dynamics resulting in the effective suppression of the influence of thermal fluctuations, and thus in the decrease of the line width [57, 59, 60,70]. When the current is increased further and approaches its value for the first jump, the contribution of the second nonlinear oscillation mode (which will dominate the spectrum after the first frequency jump) becomes visible, leading to line broadening and a decrease of the maximal spectral amplitude (see spectra for $a_J$ = 0.34, 0.36).

After the first jump, the amplitude of magnetization precession becomes large ($S_{int}$>0.3) and the relative influence of thermal fluctuations on the motion of magnetization becomes small. For this reason, the line width for most spectra of the second group (except for those close to the second frequency jump) is extremely small.

In fact, it is below the resolution limit of our simulations ($\Delta f_{\min} \approx 10$ MHz ), thus being in a good agreement with experimental observations. When approaching the current value of the second frequency jump, the line width starts to increase again (and the maximal spectral power decreases) due to the influence of the next nonlinear mode.

For the last spectral group, the line width and the maximal value of the spectral power exhibit the same non-monotonic behavior. However, the line broadening for the large current values ($a_J > 0.90$) in this region is not due to the next incipient frequency jump but due to the onset of spatially incoherent magnetization dynamics (note that the maximal experimentally used current value $I_{\max} = 10$ mA corresponds to $a_J \approx 1.0$). The line broadening for I > 9.0 mA is also observed experimentally and is clearly visible in Fig. 2(c). However, for values of $a_J$ greater than the second frequency jump, the width of the simulated spectral peaks is substantially larger than the width of spectral lines measured for corresponding current values (computed as $I = a_J / \kappa$). Another important difference between experiment and the simulation results is that the narrowest spectral lines found in the simulations exist between the first and the second frequency jumps, while the narrowest lines observed experimentally occur after the second frequency jump.

Before proceeding to a direct comparison with the experimental oscillation frequencies and amplitudes, we note an important difference between the magnetization dynamics of the Py elliptical nanomagnet simulated in this paper and that of the Co elliptical nanoelement studied in detail previously in ref. [64]. For the Co element in [64] the spatial coherence of the magnetization oscillations was lost already for currents very close to the onset of the steady-state oscillations, followed by a transition to a completely chaotic regime [71]. In contrast to this behavior, magnetization dynamics of the Py element studied here remains nearly coherent up to current values *several times larger* than the critical current. This difference *cannot* be attributed to much lower temperatures for which the experiment discussed here has been performed (compared to room temperatures used in [10]), because the transition to the chaotic regime slightly above $a_{cr}$ was observed in [64] already for simulations performed at $T = 0$. The difference can also not be due to a slightly higher element thickness used here ($h_{Py} = 4$ nm compared to $h_{Co} = 3$ nm in [64]), because the much higher exchange constant of Co ($A_{Co} = 3 \times 10^{-6}$ erg/cm, see [64] for details) when compared with the Py exchange ($A_{Py} = 1.3 \times 10^{-6}$ erg/cm) should at least compensate this slightly larger thickness of the Py nanoelement.

We argue that this important discrepancy in the behavior of the two quite similar systems studied here and in [64] is due to the very different character of the nonlinear magnetization oscillation modes of these nanoelements. Whereas in [64] *several* oscillation modes with a quite complicated localization patterns arose and coexisted when the oscillation amplitude increased (see spatial maps in figures 1, 3 and 4 in [64]), in this work we have found that for each given current value there is a *single* nonlinear eigenmode where the oscillating spins are confined in a localized area of the nanomagnet without any node lines between these oscillating spins. It seems plausible that the transition to a quasichaotic behavior from a single mode would be inhibited compared to the case of several coexisting modes with different spatial profiles. We believe that the physical reason for excitation of a single mode in the case of substantially non-collinear magnetizations of the pinned and the free layers is that spin torque is nearly spatially uniform. Indeed, in the case of nominally collinear magnetizations, the direction and magnitude of spin torque exerted on the free layer exhibits strong spatial variations due to spatially non-uniform magnetization direction predicted by micromagnetics. This results in local magnetizations of the free and the pinned layers making small negative angles with respect to each other in some parts of the sample and small positive angles in other parts of the sample. Since spin torque is proportional to the small angle between magnetizations of the free and the pinned layers, the case of nominally collinear magnetizations gives rise to strongly spatially non-uniform spin torque. In the case of non-collinear magnetizations, small variations of the magnetization direction over the sample area result in small deviations of spin torque direction and magnitude from their average values. A spatially non-uniform pattern of spin torque is more likely to couple to multiple oscillation modes of the nanomagnet. In the case of nearly constant uniform torque, the coupling to the longest-wavelength mode is expected to be the strongest.

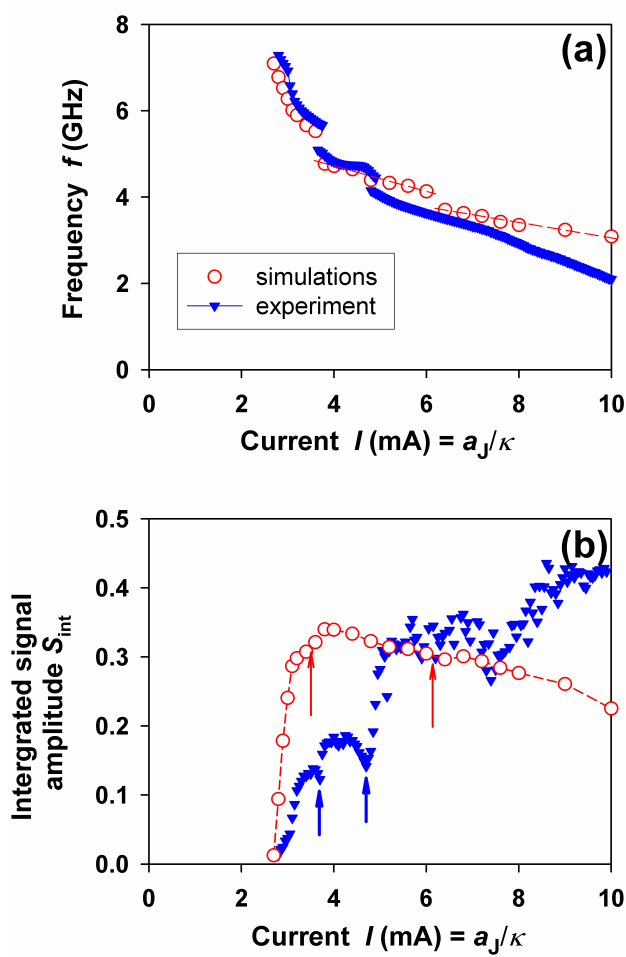

Figure 6. (Color online) (a) Comparison of the experimentally measured (solid triangles) and simulated (open circles) dependence of the frequency of oscillations on current. (b) Experimentally measured (solid triangles) and simulated (open circles) integrated rms-amplitude spectral density, $S_{int}$, as a function of current.

Figure 6 presents a direct comparison between experimental data and results of LLGS simulations. First, we show in Fig. 6(a) the current dependence of the magnetization oscillation frequency as measured experimentally (solid triangles) and as obtained from micromagnetic simulations (open circles). For plotting the simulation data as $f(I)$ we have used the conversion from the spin torque amplitude $a_J$ to the current strength $I$ in the form $I = a_J / \kappa$ with the conversion factor $\kappa \approx 0.1$ computed as explained above. The simulations reproduce the current dependence of the oscillation frequency fairly well, except for the position of the second frequency jump, which occurs in simulations at the current about 20% higher than in the experiment. However, taking into account that a nanomagnet with perfect edges was simulated and that the simulations did not contain any adjustable parameters (except the conversion factor $\kappa$) the agreement between simulations and experiment can be considered as very satisfactory, as far as the oscillation frequency is concerned.

Despite the good agreement between experiment and simulations for the dependence of the oscillation *frequency* on current, the simulations could not closely reproduce the corresponding dependence of the oscillation *amplitude* on current. Figure 6(b) shows the experimentally measured (solid triangles) and simulated (open circles) integrated signal amplitudes $S_{int}$ as functions of the bias current. The general trend of the measured oscillation amplitude is to increase gradually with increasing current, together with a series of dips at currents corresponding to the frequency jumps shown in Fig. 2(c). In contrast, the LLGS simulations predict a rapid increase of the oscillation amplitude just above the critical current for the onset of oscillations, followed by a slow gradual decrease. Some minor anomalies on the simulated $S_{int}(I)$ around the frequency jumps can be seen, however, they are far less pronounced than the corresponding experimentally-observed dips.

Taking into account the good agreement between simulations and experiment for the oscillation frequency, the discrepancy for the amplitude of oscillations is very surprising and requires a detailed analysis. The failure of our LLGS simulations to predict the correct dependence of the oscillation amplitude on current indicates that the standard micromagnetic LLGS approach for spin-torque driven excitations in nanomagnets requires modifications. Below we propose some possible routes towards improvement of the theoretical description of spin-torque-driven excitations in nanomagnets.

One possible way of solving this problem would be to introduce a nonlinear dissipation (a dependence of the dissipation parameter $\lambda$ on the rate of the magnetization change $d\mathbf{m}/dt$ in the form $\lambda = \lambda_0 (1 + q_1 \cdot (d\mathbf{m}/dt)^2 + ...)$) as suggested in [40]. In making such an attempt, one should keep in mind that a too strong nonlinearity (large values of the nonlinear coefficient $q_1$) would destroy the good agreement between simulated and measured oscillation frequencies, especially for the initial part of the $f(a_J)$-dependence where the transition between linear and nonlinear oscillation regimes is observed. However, a moderate nonlinearity could weakly affect the oscillation frequency for small to moderate oscillation amplitudes (small $a_J$), while improving the spatial coherence of the magnetization oscillations for large currents. (If the dissipation coefficient $\lambda$ increases with increasing $d\mathbf{m}/dt$, then it should strongly suppress the short-wavelength excitations which lead to incoherent magnetization oscillations). In this way, one would obtain higher oscillation powers and narrower linewidths for larger currents, thus improving the agreement between theory and experiment. Clearly, this subject requires further investigation.

Another possible way of reconciling theory and experiment for the current dependence of both frequency and amplitude of the excited modes would be the generation of spin wave modes that are more spatially non-uniform than those shown in Fig. 3(a). Indeed, if only a part of magnetization of the nanomagnet moves with large amplitude (e.g. edge modes), both a significant nonlinear shift of frequency and a relatively small average measured amplitude will be observed. Furthermore, the growth of the average amplitude of such non-uniform spin wave modes is likely to proceed via a gradual spatial growth of the oscillating domain, which should give rise to a gradual increase of the measured amplitude and result in a dependence of the amplitude on current similar to the experimentally observed dependence shown in Fig. 6(b). A possible mechanism leading to excitation of strongly spatially non-uniform modes is the instability of magnetization arising from lateral spin transport in spin valve structures [72-74]. A theoretical test of this scenario requires the development of a micromagnetic code that explicitly treats magnetization dynamics coupled to spatially non-uniform spin-dependent electrical transport, which is beyond the scope of this work. Softening of spin-wave spectrum by spin-polarized current [75,76] could also be an important factor to be taken into account for reconciling theory of current-driven excitations with the experimental results presented in this paper.

## V. CONCLUSIONS

In conclusion, we have measured the spectral properties of current-driven excitations in nanoscale spin valves with non-collinear magnetizations of the free and pinned ferromagnetic layers. We find that spin-polarized current in these devices excites a few coherent large-amplitude nonlinear modes of magnetization oscillation in the free layer. Different modes are excited in different current intervals. We find that the amplitude and the coherence of the current-driven excitations decrease in the current intervals where transitions between these modes take place. We simulate the response of magnetization to spin-polarized current in our samples by employing LLG micromagnetic simulations with a Slonczewski spin torque term [46]. These LLGS simulations capture a number of features of the experimental data: (i) the decrease of frequency of the excited oscillation modes with increasing current, (ii) downward jumps of the frequency of excitations with increasing current resulting from transitions between different oscillation modes, (iii) the high degree of coherence (narrow spectral line width) of the excited modes. However, the LLGS simulations give qualitatively incorrect predictions for the amplitude of the excited modes as a function of current. Simulations predict rapid growth of the oscillation amplitude above the threshold current for the onset of spin wave excitations, followed by a slow decrease of the amplitude. This is in sharp contrast to the more gradual increase of the oscillation amplitude with current observed in our experiment. Our results demonstrate that additional factors possibly including nonlinear damping and/or lateral spin transport need to be taken into account for a quantitative description of large-amplitude magnetization dynamics driven by spin-polarized current in magnetic nanostructures.

ACKNOWLEDGMENTS: The authors thank J. Miltat, D. Mills and A. Slavin for many useful discussions. This research was supported by the Deutsche Forschungsgemeinschaft (DFG grant BE 2464/4-1), the Office of Naval Research, and the National Science Foundation's Nanoscale Science and Engineering Centers program through the Cornell Center for Nanoscale Systems. We also acknowledge NSF support through use of the Cornell Nanoscale Facility node of the National Nanofabrication Infrastructure Network and the use of the facilities of the Cornell Center for Materials Research.